\documentclass[aps,prl,twocolumn,showpacs,superscriptaddress]{revtex4}
\usepackage{graphicx}

\begin{document}


\title{Strong Upper Limits on Sterile Neutrino Warm Dark Matter}

\author{Hasan Y{\"u}ksel}
\affiliation{Department of Physics, Ohio State University,
Columbus, Ohio 43210}
\affiliation{Center for Cosmology and Astro-Particle Physics,
Ohio State University, Columbus, Ohio 43210}

\author{John F. Beacom}
\affiliation{Department of Physics, Ohio State University,
Columbus, Ohio 43210}
\affiliation{Center for Cosmology and Astro-Particle Physics,
Ohio State University, Columbus, Ohio 43210}
\affiliation{Department of Astronomy, Ohio State University,
Columbus, Ohio 43210}

\author{Casey R. Watson}
\affiliation{Department of Physics and Astronomy,
Millikin University, Decatur, Illinois 62522}

\date{27 June 2007}

\begin{abstract}
Sterile neutrinos are attractive dark matter candidates.  Their parameter space of mass and mixing angle has not yet been fully tested despite intensive efforts that exploit their gravitational clustering properties and radiative decays.  We use the limits on gamma-ray line emission from the Galactic Center region obtained with the SPI spectrometer on the INTEGRAL satellite to set new constraints, which improve on the earlier bounds on mixing by more than two orders of magnitude, and thus strongly restrict a wide and interesting range of models. 
\end{abstract}

\pacs{95.35.+d, 13.35.Hb, 14.60.St, 14.60.Pq}


\maketitle
\vspace*{-0.2cm}

\begin{figure}[b]
\includegraphics[width=3.25in,clip=true]{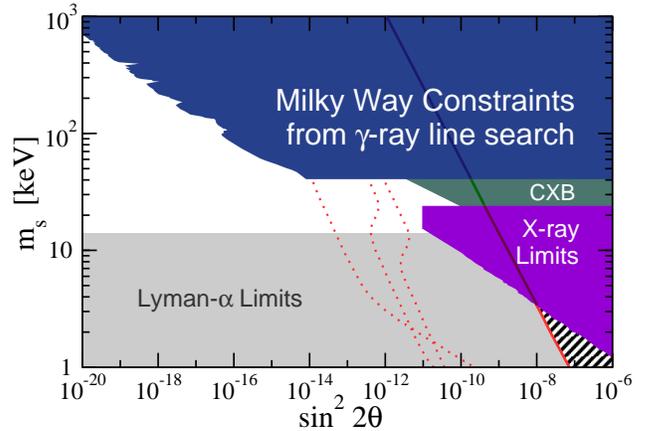}
\caption{The sterile neutrino dark matter mass $m_s$ and mixing $\sin^2 2\theta$ parameter space, with shaded regions excluded.  The strongest radiative decay bounds are shown, labeled as Milky Way (this paper), CXB~\cite{Boyarsky-CXB}, and X-ray Limits (summarized using Ref.~\cite{Watson}; the others~\cite{xraylimits} are comparable).  The strongest cosmological bounds~\cite{lymanalpha} are shown by the horizontal band (see caveats in the text).  The excluded Dodelson-Widrow~\cite{Dodelson}  model is shown by the solid line; rightward, the dark matter density is too high (stripes).  The dotted lines are models from Ref.~\cite{extralines}, now truncated by our constraints.
}
\label{fig:integral}
\end{figure}

{\bf Introduction.---}
The existence of dark matter is certain, but the properties of the dark matter particles are only poorly constrained, with several attractive but rather different candidates.  One of these, sterile neutrinos, would be a very plausible addition to the Standard Model~\cite{SMextension, Dodelson, meso, DMtheory}.  If their masses were in the range $\sim 0.1-100$ keV, they would also act as ``warm" dark matter~\cite{Dodelson, meso, DMtheory}, which could be in better accord with observations than standard ``cold" dark matter candidates.  Even if sterile neutrinos are not a dominant component of the dark matter, they may still exist and cause other interesting effects~\cite{astro, Bezrukov}, such as pulsar kicks~\cite{pulsar}, and may affect reionization~\cite{reionization}.  It is therefore important to deeply probe the sterile neutrino parameter space, as defined by the mass $m_s$ and mixing $\sin^2 2\theta$ with ordinary active neutrinos, and shown in Fig.~\ref{fig:integral}.

One means of testing sterile neutrino dark matter models is through cosmological searches, which rely on  the effects of sterile neutrino dark matter on the large-scale structure of gravitationally-collapsed objects.  While recent results based on the clustering of the Lyman-$\alpha$ forest and on other data have been interpreted as lower limits on the sterile neutrino mass of up to about 10--13 keV, independent of the mixing angle~\cite{lymanalpha}, these constraints may be weakened depending on the sterile neutrino production model (e.g., Ref.~\cite{higgssinglet}).
 
Another means of constraining sterile neutrino dark matter is through their radiative decay to active neutrinos, $\nu_s \rightarrow \nu_a + \gamma$.  These decays produce mono-energetic photons with $E_\gamma = m_s/2$.  While the decay rate is exceedingly slow due to the tiny active-sterile mixing, modern satellite experiments can detect even these very small x-ray/gamma-ray fluxes, and such a signal could specifically identify a sterile neutrino dark matter candidate.  The signal from nearby dark matter halos is line emission and the cosmic signal from all distant halos is broadened in energy by the integration over redshift.  There are limits obtained using the Cosmic X-ray Background (CXB) data~\cite{meso, Boyarsky-CXB} and, at lower masses, stronger limits using data from a variety of nearby sources (see e.g., Refs.~\cite{Watson, xraylimits} and references therein).

It is important to improve on both the cosmological and radiative decay constraints; despite intensive efforts, viable models that match the observed dark matter density still remain.  In fact, it has recently been emphasized~\cite{meso, DMtheory} that some models may extend to regions of the parameter space far from the earliest and simplest models~\cite{Dodelson} to much smaller mixing angles.  We calculate the gamma-ray flux from dark matter decays around the Milky Way center and compare this to the limits on the line emission flux from the INTEGRAL satellite.  The high sensitivity and spectral resolution of the available data enable us to derive new and very stringent constraints.  For masses above 40 keV, this improves on the CXB constraints~\cite{Boyarsky-CXB} on the mixing angle by more than two orders of magnitude.


{\bf INTEGRAL Gamma-Ray Line Search.---}
Teegarden and Watanabe have reported results from a search for gamma-ray line emission from point and diffuse sources in the Milky Way~\cite{Teegarden}, using the SPI spectrometer on the INTEGRAL satellite~\cite{INTEGRAL}.  In the energy range 20--8000 keV, they tested for lines of intrinsic width 0, 10, 100, and 1000 keV. The additional line width due to instrumental resolution increases over the above energy range from $\sim 2$ to 8 keV full-width half-maximum (FWHM).  As expected, their analysis recovered the known astrophysical diffuse line fluxes at 511~\cite{511line} and 1809 keV~\cite{1809line}, and no others, validating their procedures~\cite{Teegarden}. The principal advantages of the SPI instrument for a sterile neutrino decay search are its wide field of view and excellent energy resolution.  For sterile neutrino decays in the Milky Way halo, the line width due to virial motion is $\sim 10^{-3}$, which is therefore small enough to be neglected.

Two large-scale regions around the Galactic Center were considered, with angular radii of $13^\circ$ and $30^\circ$, and exposures of 1.9 and 3.6 Ms, respectively.  The 24$^\circ$ collimated field of view was used without the coded mask image reconstruction and the corresponding limits on the flux from an unknown line emission were derived by deconvolving an assumed sky brightness distribution (either a Gaussian with $10^\circ$ FWHM for the former or flat for the latter region) and the wide angular response of the collimator.  To improve the sensitivity to line emission specifically from these regions, the average flux away from the Galactic Center Region (angular radii of $> 30^\circ$) was subtracted from the flux from inside the Galactic Center Region. This procedure cancels almost all of the instrumental backgrounds. This also cancels all of the cosmic signal and part of the halo signal, and a careful calculation of the latter effect is taken into account in our analysis.  For the Galactic Center Region, the $3.5\sigma$ limits on narrow line emission are $\lesssim 10^{-4}$ photons cm$^{-2}$ s$^{-1}$ for the full range of energies. The actual energy dependence of the limiting flux, ${\cal F}_{lim}(E)$, is more complicated, and we took this into account (leading to the slightly jagged edge of our exclusion region).


{\bf Milky Way Dark Matter Decay Flux.---}
To turn the INTEGRAL limits on generic line emission into constraints on sterile neutrino dark matter, we calculated the expected gamma-ray emission from the decay of sterile neutrinos in the Milky Way (the INTEGRAL limits also strongly constrain certain decays of GeV-mass dark matter models~\cite{Yuksel:2007dr}).  For a long-lived decaying sterile neutrino with lifetime $\tau$ and mass density $\rho = m_s n$, the intensity~\cite{SteckerBook} (number flux per solid angle) of the decay photons coming from an angle $\psi$ relative to the Galactic Center direction is
\begin{equation}
{\cal I}(\psi) = \frac{\rho_{sc} R_{sc}} {4\pi m_s \tau} {\cal J}(\psi)\,,
\label{eq:intensity}
\end{equation}
where the dimensionless line of sight integral,
\begin{equation}
{\cal J}(\psi) = \frac{1}{\rho_{sc} R_{sc}} \int_{0}^{\ell_{max}} d\ell \; 
\rho\left(\sqrt{R_{sc}^2-2\, \ell\, R_{sc}\cos\psi+\ell^2} \right)\,,
\label{eq:Jintegral}
\end{equation}
is normalized at the solar circle, with $R_{sc} = 8.5$~kpc and $\rho_{sc} = 0.3$ GeV cm$^{-3}$ (these cancel later).  While $\ell_{max}$ depends on the adopted size of the halo, contributions beyond the scale radius of the density profile, typically about 20--30 kpc, are negligible.

The sterile neutrino radiative lifetime $\tau$ is
\begin{equation}
\frac{1}{\tau} = \left(6.8 \times
10^{-33} {\rm\ s}^{-1}\right) 
\left[ \frac{\sin^2 2\theta}{10^{-10}}\right] \left[\frac{m_s}{\rm keV}\right]^5,
\end{equation}
where we have chosen the Dirac neutrino decay lifetime~\cite{lifetime}; for the Majorana case, which may be favored, the lifetime is 2 times shorter, which would lead to more restrictive constraints. The prefactor in Eq.~(\ref{eq:intensity}) can then be expressed in terms of the mass and mixing of the sterile neutrino,
\begin{equation}
\frac{\rho_{sc} R_{sc}} {4\pi m_s \tau} = 
\left(4.3 \times 10^{-6} {\rm\ cm}^{-2} {\rm\ s}^{-1} {\rm\ sr}^{-1}\right)
\left[ \frac{\sin^2 2\theta}{10^{-10}}\right] \left[\frac{m_s}{\rm keV}\right]^4\,.
\label{eq:propfac}
\end{equation}
The number flux of photons at energy $E_\gamma=m_s/2$ is obtained by integrating the intensity, Eq.~(\ref{eq:intensity}), over the field of view,
\begin{equation}
{\cal F}_s=  \int_{\Delta \Omega}  d\Omega \; {\cal I}(\psi)
= \frac{\rho_{sc} R_{sc}} {4\pi m_s \tau} \int_{\Delta \Omega}  d\Omega\; {\cal J}(\psi)\,,
\label{eq:decflux}
\end{equation}
where the solid angle is $ \Delta \Omega = 2 \pi (1 - \cos \psi)$.

The dark matter distribution of the Milky Way is not perfectly known~\cite{Klypin:2001xu}, though the variations between models make little difference for dark matter decay, since the density appears only linearly in the calculations (unlike for dark matter annihilation, where it appears quadratically).  A trivial lower bound for the integral in Eq.~(\ref{eq:decflux}) can be obtained by taking the dark matter density to be constant within some radius from the Galactic Center, which we take to be $R_{sc}$.  Then the line of sight and field of view integrals are just multiplications: using Eq.~(\ref{eq:Jintegral}), the former is $\simeq 2$, and since $\Delta \Omega \simeq 0.16$ for $\psi = 13^\circ$, the latter is $\int_{\Delta \Omega} d\Omega\; {\cal J}(\psi) \simeq 0.3$. 

\begin{figure}[b]
\includegraphics[width=3.25in,clip=true]{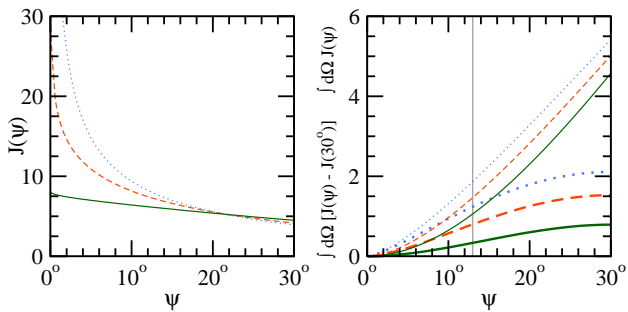}
\caption{{\it Left panel:} The line of sight integral ${\cal J}(\psi)$ as a function of the pointing angle $\psi$ with respect to the Galactic Center direction for the three different profiles considered (Kravtsov, NFW, and Moore, in order of solid, dashed and dotted lines).  {\it Right panel:}  Integrals up to the angle $\psi$ of ${\cal J}(\psi)$ (thin upper lines) and ${\cal J}(\psi) - {\cal J}(30^{\circ})$ (thick lower lines).  The grey line at $13^\circ$ marks  the field of view for the INTEGRAL flux limit, and we chose $\int_{\Delta\Omega} d\Omega \, [{\cal J}(\psi) - {\cal J}(30^{\circ})] \simeq 0.5$ as a conservative value for our subsequent constraints.}
\label{fig:los}
\end{figure}

For realistic dark matter density profiles, the field of view integral in Eq.~(\ref{eq:decflux}) will be larger, since the density is larger (though more uncertain) in the central region.  We calculated this for the Navarro-Frenk-White (NFW)~\cite{Navarro:1995iw}, Moore~\cite{Moore:1999gc}, and Kravtsov~\cite{Kravtsov:1997dp} profiles, which are all commonly used (see also Ref.~\cite{othprofiles}).  These are normalized with $\rho(R_{sc}) = $ 0.30, 0.27, and 0.37 GeV cm$^{-3}$, respectively.  These slight differences in normalization compensate the different slopes at inner radii so that the masses enclosed at outer radii are the same~\cite{Klypin:2001xu}.  In the left panel of Fig.~\ref{fig:los}, the thin lines show ${\cal J}(\psi)$ as a function of the angle $\psi$ for each profile; in the right panel, the corresponding thin lines show these integrated over the field of view (up to the angle $\psi$), as in Eq.~(\ref{eq:decflux}).  These results take into account the variation of density with position, and also the contribution from halo dark matter beyond the solar circle on the other side of the Milky Way.  Note that all three profiles have similar values of $\int_{\Delta \Omega} d\Omega\; {\cal J}(\psi)$, since the large field of view de-emphasizes the inner radii where the differences between the profiles are the largest.

{\bf Constraints on Sterile Neutrinos.---}
As noted above, the INTEGRAL limits on line emission from the Galactic Center region are obtained by subtracting the average flux outside this region ($\psi > 30^\circ$) from the flux inside this region ($\psi < 13^\circ$), which must be taken into account in our analysis.  
To be conservative, we considered the maximum effect of this subtraction by fixing the intensity outside the Galactic Center region to its value at $\psi = 30^\circ$. (In fact, it is smaller at larger angles.)  In terms of our equations, this is
\begin{equation}
\Delta{\cal F}_s=\frac{\rho_{sc} R_{sc}} {4\pi m_s \tau} \int_{\Delta \Omega}  d\Omega \,
 \left[{\cal J} (\psi)-{\cal J}(30^\circ)\right]\,.
\label{eq:deltaF}
\end{equation}
In the right panel of Fig.~\ref{fig:los}, our results for the integrated ${\cal J}(\psi)- {\cal J}(30^{\circ})$ are shown by the thick lines.  The effect of this subtraction correction is not large, less than a factor of 3 at $\psi = 13^\circ$ for all three profiles. 
In addition, the INTEGRAL flux limits of Ref.~\cite{Teegarden} for an angular region of $\psi < 13^\circ$ assume that the line emission intensity follows a two-dimensional Gaussian with FWHM of $10^\circ$, while a flat-source profile would yield somewhat weaker limits. 
 
To shield our results from such uncertainties associated with the distribution of dark matter in the Milky Way, including whether warm dark matter profiles are less centrally concentrated than cold dark matter profiles, we use a rather conservative value, $\int_{\Delta\Omega} d\Omega \, [{\cal J}(\psi) - {\cal J}(30^{\circ})] \simeq 0.5$, in our subsequent calculations.
Our results can be easily rescaled for a different value and our limits should improve as the amount of data increases in time.

While we have presented our results for the region within $13^\circ$ of the Galactic Center, there are also flux limits for an angular region of $\psi < 30^\circ$ and an assumption that the intensity is constant in angle~\cite{Teegarden}.  The flux limits for $\psi < 30^\circ$ are $\simeq 3$ times weaker than those for $\psi < 13^\circ$~\cite{Teegarden}.  However, as shown in the right panel of Fig.~\ref{fig:los}, the sterile neutrino decay flux, which is proportional to $\int_{\Delta\Omega} d\Omega \, [{\cal J}(\psi) - {\cal J}(30^{\circ})]$, is $\simeq$ 2--3 times larger for $\psi < 30^\circ$ than for $\psi < 13^\circ$,  compensating the lower sensitivity.  Thus our results are rather robust against the choice of angular region used and other assumptions
for analyzing the INTEGRAL limits.

With these detailed results on the sterile neutrino dark matter distribution, we define constraints in the parameter space of mass and mixing.  The expected line flux at $E_\gamma = m_s/2$ from dark matter decay, which depends on $m_s$ and $\sin^2 2\theta$, should not exceed the INTEGRAL limits (for $3.5\sigma$), i.e., ${\cal F}_{lim} > \Delta{\cal F}_s$, or
\begin{equation}
{\cal F}_{lim}(E) > 
\frac{\rho_{sc} R_{sc}} {4\pi m_s \tau}
\int_{\Delta\Omega} d\Omega \, [{\cal J}(\psi)-{\cal J}(30^{\circ})]\,.
\label{eq:constraint}
\end{equation}

Substituting  Eq.~(\ref{eq:propfac}) and $\int_{\Delta\Omega} d\Omega \, [{\cal J}(\psi) - {\cal J}(30^{\circ})] \simeq 0.5$ yields our result in Fig.~\ref{fig:integral}.  The boundary of the excluded region is jagged on the left due to the actual energy dependence of the limiting flux, ${\cal F}_{lim}(E)$  (see Fig.~9 of Ref.~\cite{Teegarden}).  The energy range available with the SPI instrument causes the sharp cut-off at $m_s = $ 40 keV.  Our constraint is coincidentally in line with prior constraints at lower masses using the x-ray emission from nearby sources.  There is only a narrow gap, $m_s \simeq $ 20--40 keV, in which the best available mixing constraints are substantially weaker.  The constraints shown in Fig.~\ref{fig:integral} assume that sterile neutrinos comprise all of the required present-day dark matter, but the limits at large mass are so stringent that they would provide strong limits even on sterile neutrinos that were only a fraction of the dark matter.


{\bf Conclusions.---}
Sterile neutrinos require only a minimal and plausible extension of the Standard Model~\cite{SMextension, Dodelson, meso, DMtheory} and can solve problems in reconciling the observations and predictions of large-scale structure~\cite{meso, DMtheory}.   Despite intensive efforts on setting constraints, there are still viable sterile neutrino  dark matter models over a wide range of mass $m_s$ and mixing $\sin^2 2\theta$; the focus is now at larger mass and smaller mixing than considered in the earliest and simplest models~\cite{Dodelson}.  In this region, the models are very challenging to test, either through their differences in clustering with respect to cold dark matter candidates~\cite{lymanalpha} or their astrophysical effects~\cite{pulsar, reionization, astro}), or through their very small radiative decay rates~\cite{meso, Boyarsky-CXB, Watson, xraylimits} or laboratory tests~\cite{Bezrukov}. 
  
Teegarden and Watanabe~\cite{Teegarden} presented the results of a sensitive search for line emission in the Galactic Center Region, using data from the SPI spectrometer on the INTEGRAL satellite~\cite{INTEGRAL}.  Based on a simple and conservative calculation of the expected gamma-ray flux from sterile neutrino dark matter decays, we have used these limits to set new and very strong constraints on sterile neutrino parameters, as shown in Fig.~\ref{fig:integral}.  The large-mass region is now very strongly excluded, improving on the previous CXB mixing constraints~\cite{Boyarsky-CXB} by more than two orders of magnitude.  At fixed $m_s$, the boundary in $\sin^2 2\theta$ is defined by the $3.5\sigma$ exclusion; using Eqs.~(\ref{eq:constraint}) and (\ref{eq:propfac}), it is easy to see that points with $\sin^2 2\theta$ values ten times larger than at the boundary are excluded by a nominal $35\sigma$, and so on.  On the scale of the figure, any reasonable further degradations in the conservatively-chosen inputs would not be visible.  We anticipate that it will be possible to extend our constraints, in particular going to lower masses, by dedicated analyses of the INTEGRAL data, which we strongly encourage.  If the sensitivity of this and other techniques can be improved upon, then it may be possible to definitively test sterile neutrinos as a dark matter candidate.


\medskip
We thank  Matt Kistler, Bonnard Teegarden, and Ken Watanabe for helpful comments.
HY and JFB were supported by NSF CAREER Grant PHY-0547102
to JFB.


\end{document}